\documentclass[aps,twocolumn,nopacs,floats,superscriptaddress]{revtex4}

\usepackage[colorlinks=true, citecolor=blue, linkcolor=blue, urlcolor=blue]{hyperref}
\usepackage{graphicx,dcolumn,longtable,epsfig}
\usepackage[usenames]{color}
\usepackage{amssymb}
\usepackage{amsmath}
\usepackage{bm}
\usepackage{footnote}
\usepackage{float}
\usepackage{subfigure}
\usepackage{color}

\usepackage{ulem}
\usepackage[T1]{fontenc}

\newcommand{\be}{\begin{equation}}
\newcommand{\ee}{\end{equation}}

\def\bea{\begin{eqnarray}}
\def\eea{\end{eqnarray}}

\begin{document}
%\title{Critical Temperature of Coulomb Bose Gases in Two and Three Dimensions}
\title {Superconducting transition temperature of the Bose one-component plasma}

\author{Chao Zhang}
\affiliation{Department of Modern Physics, University of Science and Technology of China, Hefei, Anhui 230026, China}
\affiliation{Hefei National Laboratory, University of Science and Technology of China, Hefei 230088, China}

\author{Barbara Capogrosso-Sansone}
\affiliation{Department of Physics, Clark University, Worcester, Massachusetts 01610, USA}

\author{Massimo Boninsegni}
\affiliation{Department of Physics, University of Alberta, Edmonton, Alberta, Canada T6G 2H5}

\author{Nikolay V. Prokof'ev}
\affiliation{Department of Physics, University of Massachusetts, Amherst, MA 01003, USA}

\author{Boris V. Svistunov}
\affiliation{Department of Physics, University of Massachusetts, Amherst, MA 01003, USA}
\affiliation{Wilczek Quantum Center, School of Physics and Astronomy, Shanghai Jiao Tong University, Shanghai 200240, China}

\begin{abstract}
We present results of numerically exact simulations of the Bose one-component plasma, i.e., a Bose gas with pairwise Coulomb interactions among particles and a uniform neutralizing background.
We compute the superconducting transition temperature 
for a wide range of densities, in two and three dimensions, for  both continuous and lattice versions of the model. 
%MB
The Coulomb potential causes the weakly interacting limit to be approached at high density, but gives rise to no qualitatively different behavior, {\it vis-a-vis} the superfluid transition, with respect to short-ranged interactions.
%MB
Our results are of direct relevance to quantitative studies of  
 bipolaron mechanisms of (high-temperature) superconductivity.
\end{abstract}

%\pacs{}
\maketitle

{\it Introduction}. 
%MB Generically, the problem of 
%MB {\it ab initio} estimation of 
Reliably estimating the critical temperature ($T_c$) for the superfluid/superconducting transition in a 
correlated  
Fermi system---even with relatively large uncertainties---is a difficult task. Exponential sensitivity of $T_c$ to system 
parameters leads to the challenge of accurate and unbiased treatment
of many-body correlation effects. An exception
to this  rule are dimerized states when pairing of fermions 
takes place in real, rather than momentum, space, and the system can be effectively regarded as an assembly of interacting Bose particles.

Remarkably, the value of $T_c$ in the three-dimensional (3D) 
interacting Bose system is very close to the Bose-Einstein condensation point for the ideal gas,
$T_{c}^{(0)}\approx 3.31\ (n^{2/3}/m)$, where $n$ is the number density and $m$ the particle mass (we use units in which $\hbar= k_B=1$),
when interactions are short-range. This holds all the way (within, say, a factor two) from the
dilute gas limit to interparticle distances of the order of 
the potential range
%MB added on 11/4/2022
; for the case of superfluid $^4$He, for example, it is $T_c/T_{c}^{(0)}\approx 0.7$. 
%MB.
The archetypal model displaying this 
physical behavior is a Bose fluid of particles interacting via short-ranged, hard-core potentials \cite{Gruter97,Nho,Pilati2008,Kora2020}.
\\ \indent
In two dimensions (2D) there is no Bose-Einstein condensation, but in the presence of interactions a Bose fluid undergoes a Berezinskii-Kosterlitz-Thouless (BKT) superfluid transition,
%MB
which is theoretically well understood and, for 
%, with $T_c$ vanishing in the ideal gas limit. For 
a gas with short-ranged interactions, precisely characterized quantitatively \cite{Oliver}. 
%Moreover, as long as the system's density is not too small, or, similarly, the strength of interaction is not too weak, the expression for $T_c$ is reminiscent of its 3D counterpart.Namely, 
In a broad range of density (including the regime of strong correlations \cite{Pilati2008}), it is $T_c\approx 
1.3\ T^{(0)}$ with $T^{(0)} = (n/m)$.
%MB, and this expression   
%MB can be used for rather accurate estimation of $T_c$ . 
Thus, an approximate estimate for $T_c$ is given by 
\begin{equation}
T_c \approx  C_d\ \frac{n^{2/d}}{ m },  \ {\rm with}\  C_d =  \biggl \{ 
\begin{array}{c} 1.30\ ,\ \ \ d=2 \\ 3.31\ ,\ \ \ d= 3 \end{array}
\label{Tc_est}
\end{equation}
$d$ being the dimensionality.
Eq. (\ref{Tc_est}) can also be utilized to estimate $T_c$ for a system of dimerized fermions with short-ranged interactions, upon replacing $m$ with the effective dimer mass $m^\star$.

Recently, interest in dimerized Fermi systems has been renewed by proposals of bipolaron (high-temperature) superconductivity \cite{Mona2010,Sous2018,Zhang2021,Sous2021,Zhang2022a,Zhang2022b}. It was found that bipolarons emerging in lattice models with phonon-modulated hopping feature relatively light effective 
masses and small size, making them potential candidates for a new class of high-temperature superconductors \cite{Zhang2022b}. 
At large distances bipolarons interact via the long-range  electrostatic (Coulomb) potential. So far, this aspect of the problem has not been addressed quantitatively, as only models with short-range repulsion have been studied. 

The long-range character of interactions leads to a number of fundamental questions and changes, including the reversal of 
the relation between the low-density and weak-interaction limits.
In a Coulomb system, the dimensionless coupling parameter---the Wigner-Seitz radius $r_s$---is defined as follows in 2D and 3D:
\begin{equation}
r_s  = \frac{1}{a_B \ n^{1/d}} \times  \left\{\begin{array}{c} \pi ^{-1/2} \, , \qquad  \quad d=2 \, ,\\  (4 \pi /3)^{-1/3} \, , \quad d=3 \, ,
\end{array}\right.
\label{r_s}
\end{equation}
$a_B$ being the Bohr radius.
The perturbative regime ($r_s \ll 1)$ now corresponds to the limit of high density, where the very notion of compact bipolarons may become ill-defined, while in the low-density limit ($r_s \gg 1$) strong correlations 
may result in large deviations of $T_c$ from the ideal gas value. 
\\ \indent
%MB
On the one hand, this raises the question of applicability of Eq.~(\ref{Tc_est}) 
to realistic bipolaronic systems%MB
, calling for a reliable, quantitative verification based on unbiased, robust theoretical methods; 
%MB
on the other hand, 
the prospects of bipolaron superconductivity with large $T_c$ 
contribute to the fundamental interest in the superconductivity 
of a charged Bose gas and its broad relevance in astrophysics \cite{Ninham,Ginzburg}.

A system of charged particles in the presence of a neutralizing, uniform charge background is typically referred to as the one-component plasma (OCP). 
It has been extensively investigated in the classical case \cite{Salpeter1959,Abe1959,Brush1966,Slattery1980,Baus1980}, and there exists also a number of studies of its quantum phase diagram, for both types of quantum statistics \cite{Fetter,MA,Panat,Bishop,Ceperley1980,Ortiz1994,Moroni1995,Jones1996,Depalo2004,Clark2009}. However, to the best of our knowledge 
a systematic study of the superconducting transition for the boson case, especially in the interesting $r_s \gtrsim 1$ regime, is still missing.

%MB
In this Letter, we determine the superconducting transition critical temperature (as a function of $r_s$) in the Bose OCP, in 2D and 3D, using state-of-the-art computer simulations. 
%MB
Besides the continuous-space system, we consider a lattice model in order to gauge the effects of long-range interactions against discreteness of space and/or extra short-ranged repulsion. 
%MB
Our results are numerically ``exact'', i.e., statistical and systematic errors can be rendered negligible in practice, using ordinary computing infrastructure. 
%MB
We use the lattice \cite{worm_lattice} and continuous-space \cite{worm_cont} versions of the worm algorithm. 
\\ \indent
In order to account for the long-range character of the Coulomb interaction, we used the Ewald summation method in the pairwise form (see, for instance, Refs.~\cite {Yi2017,Wang2019}) in our lattice simulations; for simulations in continuous space we made use of the Modified Periodic Coulomb scheme of Fraser {\it et al.} \cite{Fraser1996}, affording greater computational efficiency. Both schemes are exact, when 
extrapolated to the infinite-size limit.

\begin{figure}[h]
\includegraphics[width=0.45\textwidth]{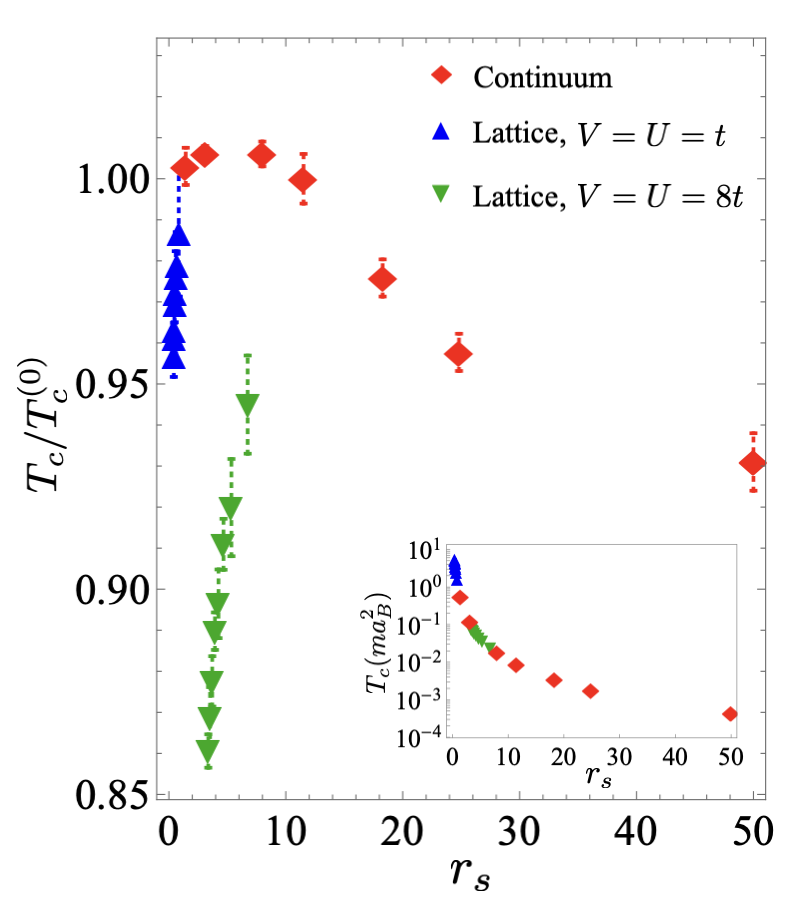} 
\caption{Transition temperature (in units of 
$T_c^{(0)}$) in 3D as a function of $r_s$, see Eq.~(\ref{r_s}). 
Blue up and green down triangles correspond to the lattice models with $V=U=t$ and $V=U=8t$, respectively. Red diamonds correspond to the continuous-space model. 
If not visible, error bars are within symbol size. 
%MB
Note that different values of $V/t$ correspond to different values of the Bohr radius $a_B$ on a lattice (see text), and this must be taken into account when comparing lattice and continuum results for the same $r_s$.
Inset: $T_c$ (in the unit of $1/(m a_B^2)$) as a function of $r_s$.
}
\label{Fig:T_c_3D}
\end{figure}

\begin{figure}[h]
\includegraphics[width=0.45\textwidth]{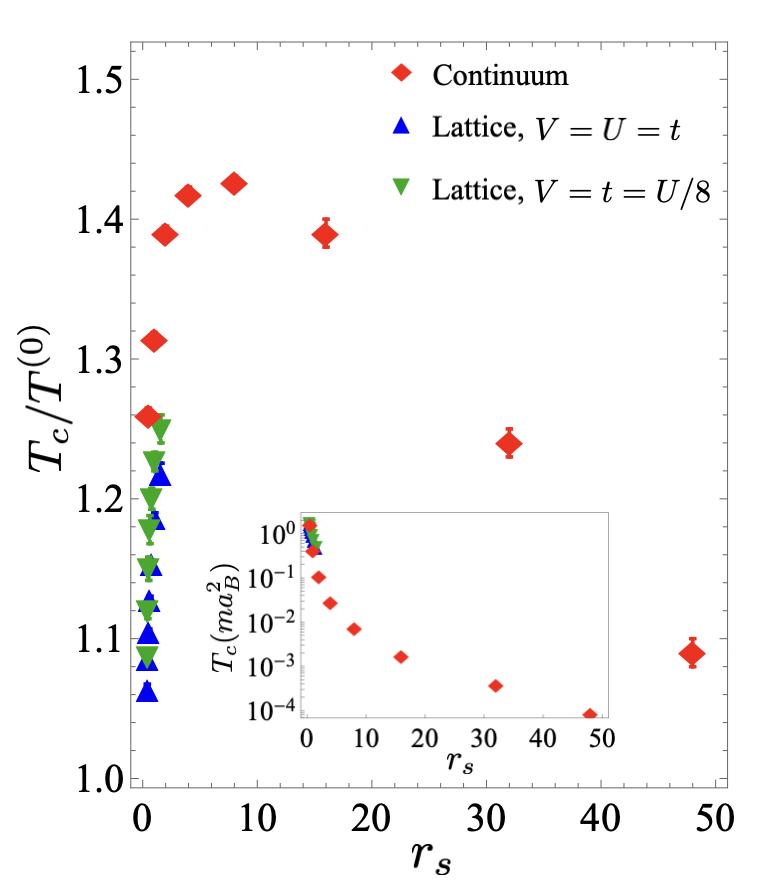}
\caption{Transition temperature  (in units of 
$T^{(0)}$) in 2D as a function of $r_s$. 
Blue up and green down triangles correspond to the lattice models with $V=U=t$ and $V=t=U/8$, respectively. Red diamonds correspond to the continuous-space model. 
Inset: $T_c$ (in the unit of $1/(ma_B^2)$) as a function of $r_s$. If not visible, error bars are within symbol size.
}
\label{Fig:T_c_2D}
\end{figure}

Our key results are presented in Figs.~\ref{Fig:T_c_3D} and \ref{Fig:T_c_2D}. 
%MB
We use $T_c^{(0)}$ ($T^{(0)}$), defined in the Introduction, as our reference temperature in 3D (2D). 
%MB
In 3D, we find the effect of Coulomb interactions to be surprisingly weak. At $r_s < 10$, the difference between $T_c$ and $T_c^{(0)}$ is within $\sim 1\%$, remaining less than  $5\%$ up to $r_s=25$. The effects of discreteness and on-site repulsion---rather small on the absolute scale---are more important than the effects of Coulomb forces. 

The situation in 2D is similar. Here the critical temperature reaches its maximum value $T_c^{\rm (max)}\approx 1.43\, T^{(0)}$ at $r_s \approx 8$ and then slowly decreases with increasing $r_s$, in a qualitative analogy with the 3D case, but with somewhat more pronounced quantitative effect of about $30 \%$. 
We traced the dependence of $T_c$ on $r_s$ up to $r_s= 48$, beyond which the superconducting, Wigner crystal, hexatic, normal metal, and emulsion phases \cite{emulsions} start competing with each other
\cite{Depalo2004,Clark2009}. As one might have expected by analogy with the known behavior of the weakly-interacting 2D Bose gas with short-range interaction, the small-$r_s$ regime is characterized by a moderate suppression of $T_c$ compared to $T_c^{\rm (max)}$. Thus, the estimate 
(\ref{Tc_est}) remains rather accurate in both 3D and 2D.
%%%%%%%%%%%%%%%%%%%%%%%%%%%%%%%%%%%%%%%%
%%%%%%%%%%%%%%%%%%%%%%%%%%%%%%%%%%%%%%
%%%%%%%%%%%%%%%%%%%%%%%%%%%%%%%%%%%%%%%
%%%%%%%%%%%%%%%%%%%%%%%%%%%%%%%%%%%%%%%

{\it The models}. The continuous-space OCP model is standard (see, for instance, Ref.~\cite{Depalo2004}). The lattice  version is described by the following Hamiltonian 
on the hypercubic lattice, with periodic boundary conditions (to ensure uniform density):
\begin{equation}
\hat H = -t \sum_{\langle {\bf i}, {\bf j} \rangle}  (\hat a_{\bf i}^{\dagger} \hat a_{\bf j}^{\,} + {\rm H.c.})+\frac{U}{2}  \sum_{\mathbf{i}} \hat n_{\mathbf{i}}^2
 +\frac{V}{2}\sum_{{\bf i} \ne {\bf j}}  
 \frac{\hat n_{\bf i} \hat n_{\bf j} }{ | {\bf i} - {\bf j} |  } \, .
\label{H_lattice}
\end{equation}
Here, the $\langle {\bf i}, {\bf j} \rangle$ sum runs over all pairs of nearest-neighbor sites indexed 
by integer vectors, $\hat a_{\bf i}^{\,}$ and $\hat n_{\bf i}= \hat a_{\bf i}^{\dagger} \hat a_{\bf i}^{\,}$ are the Bose annihilation operator and occupation number on site ${\bf i}$, respectively; $t$ is the hopping amplitude,
$U$ is the on-site (Hubbard) repulsion, and $V$ is the amplitude of the repulsive Coulomb potential.

In the model (\ref{H_lattice}), the essential---up to the choice of units---dependence of $T_c$ is three-parametric. A natural set of three dimensionless parameters consists of
$U/t$, $V/t$, and the filling factor $n=\langle n_{\bf i}\rangle$. In this work, we are not interested in the effects of commensurability (e.g., Mott physics), which model (\ref{H_lattice}) definitely demonstrates at
$n=1,\,  1/2$, and other commensurate fillings  (at strong enough interactions); hence, we work with $n \leq 0.4$.

In the $n\to 0$ limit, Eq. (\ref{H_lattice}) reproduces the single-parametric continuous-space behavior because 
$U/t$ becomes irrelevant, and $V/t$ and $n$ are absorbed 
into a single parameter $r_s$. 
The expression for $r_s$ in terms of $V/t$ and $n$ readily follows from (\ref{r_s}) by relating density and filling factor by 
$a^d$ and noting that for the tight-binding dispersion relation
$m=1/2a^2t$. Then, the amplitude of the Coulomb potential at the lattice spacing, $V=e^2/a$, immediately leads to $a_B=1/(mVa)=a(2t/V)$ (for large $V/t$ we formally have $a_B/a <<1$ implying that lattice effects are more pronounced at large $r_s$).

One question, of both fundamental and applied (see the numeric protocol below) interest, is whether the long-range Coulomb interaction changes the universality class of the superfluid transition \cite{Note}. It is easy to argue (and validate numerically)
that the transition remains in the $d$-dimensional XY universality class observed in systems with short-range interactions. Indeed, the qualitative effect of the long-range potential---incompressibility of the system---does not change properties of large-scale vortexes, which are the degrees of freedom responsible for the XY criticality  (see, e.g., \cite{our_book}). Along similar lines, recall that the 
classical XY model itself is ``incompressible'' by construction
because it lacks degrees of freedom associated with density. 

%%%%%%%%%%%%%%%%%%%%%%%%%%%%%%%%%%
\begin{figure}[t]
\includegraphics[width=0.48\textwidth]{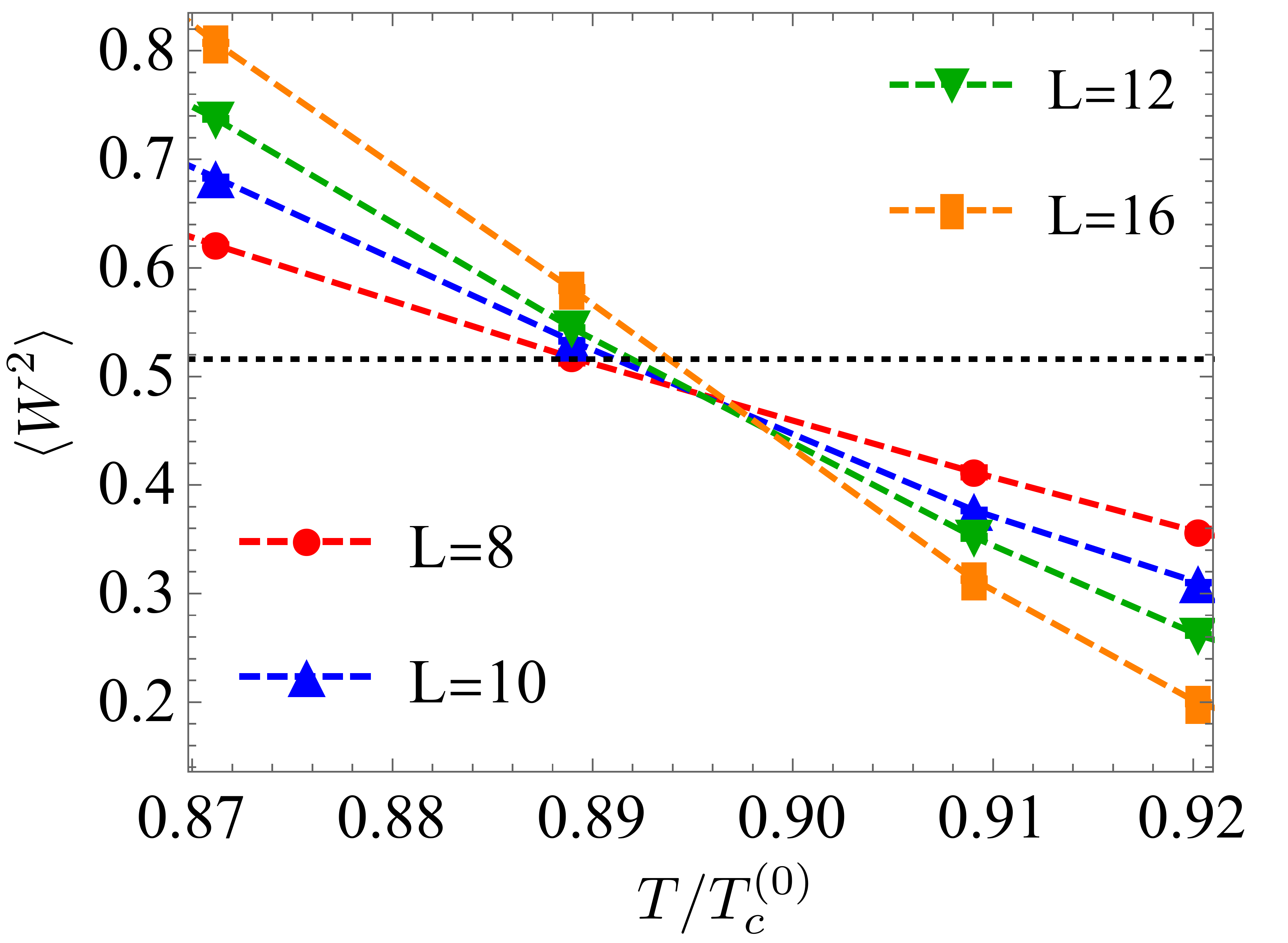}
\caption{Determining critical temperature of the 3D model (\ref{H_lattice}) from the crossing point of the mean-square winding number curves,  
$\langle W^2 \rangle$, computed for system sizes 
$L=8$, 10, 12, and 16 (red circles, blue squares, green up 
triangles, and orange down triangles, respectively)
as functions of temperature. In this example, $U=V=8t$ and $n=0.2$ (corresponding to $r_s=4.24$). 
If not visible, error bars are within symbol size. The crossing points locations pinpoint $T_c=0.896(5) T_c^{(0)}$ and their vertical values are close to the 
3D XY universality prediction of 0.516(1) \cite{universal3DXY} shown with a horizontal dashed line; small deviations from $0.516$ 
are explained by leading corrections to scaling (see the text and Fig.~\ref{Fig:T_c_L_3D}).
}
\label{Fig:W_3D}
\end{figure}
%%%%%%%%%%%%%%%%%%%%%%%%%%%%%%%%%%%%%%%%%

%%%%%%%%%%%%%%%%%%%%%%%%%%%%%%%%%%%%%%%%% 
\begin{figure}[t]
\includegraphics[width=0.5\textwidth]{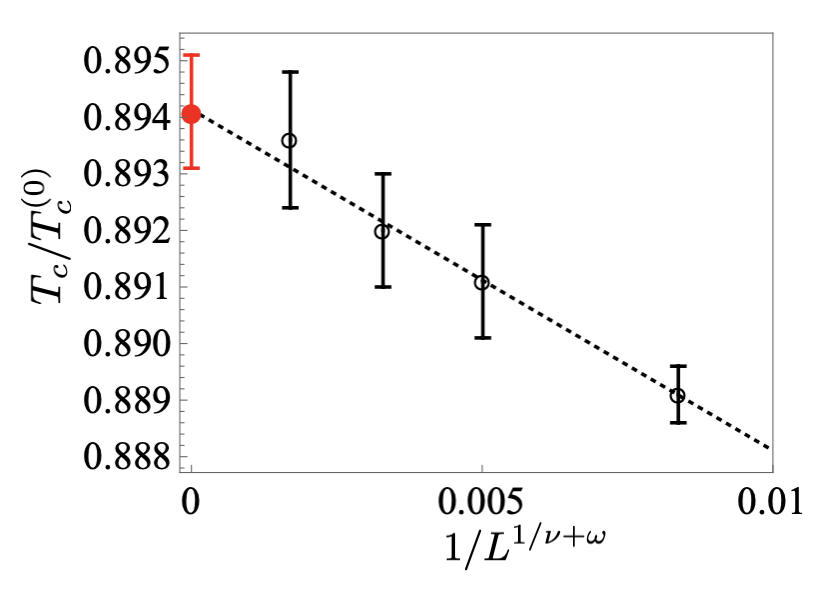}
\caption{Extracting $T_c$ from the known behavior of the leading correction to scaling at the 3D XY criticality (see the text). The system's parameters are the same as in Fig.~\ref{Fig:W_3D}. 
The extrapolated critical temperature is estimated as  
$T_c=0.8941(10) T_c^{(0)}$ (red dot).}
\label{Fig:T_c_L_3D}
\end{figure}
%%%%%%%%%%%%%%%%%%%%%%%%%%%%%%%%%%%%%%%%%%%%%%%%%

{\it Numerical protocols}.---We simulate our models with the lattice \cite{worm_lattice} and continuous-space \cite{worm_cont} versions of 
the worm algorithm. The key observable is the mean-square worldline winding number, $\langle W^2\rangle= \sum_{i=1}^d\langle W_i^2\rangle / d$, where $W_i$ is the winding number of the particle paths along the $i$-direction. 
In the thermodynamic limit  $L\to \infty$ ($L$ is the linear size
of the hypercubic simulation cell), this quantity gives access to the superfluid stiffness, $\Lambda_s$, via the Pollock--Ceperley formula \cite{Pollock_Ceperley1987}:
\begin{equation}
\Lambda_s= {\langle W^2 \rangle \, T  \over L^{d-2}} \qquad \qquad (L\to \infty)\, .
\label{PC}
\end{equation}
In the 3D case, thanks to scale invariance of the 3D XY criticality, it is more convenient to work directly with $\langle W^2 \rangle$. 
At $T_c$, this quantity saturates to the universal (system- and $L$-independent) constant $0.516(1)$ \cite{universal3DXY}, while at $T < T_c$ ($T > T_c$) it diverges (vanishes) as $L\to \infty$. These properties lead to simple and accurate schemes (illustrated in Figs.~\ref{Fig:W_3D} and \ref{Fig:T_c_L_3D}) of extracting $T_c$ and automatically validating that the universal 3D XY critical behavior does take place (from the dependence of $\langle W^2 \rangle$ on $T$ at different $L$'s). System sizes  $L \leq 16$ prove sufficient for obtaining results for $T_c$ in the model (\ref{H_lattice})
with sub-percent accuracy.

As expected, the values of $\langle W^2 \rangle$ at the crossing  points in Fig.~\ref{Fig:W_3D} are close to the universal 3D XY number. They are supposed to deviate from $0.516$ due to corrections to scaling, with the leading term vanishing as $1/L^{\omega}$ with $\omega \approx 0.8$ \cite{universal3DXY}. Taking this correction into account and obtaining   
a more accurate scheme for determining $T_c$ is achieved in two simple 
steps. First, define the ``finite-size'' critical temperature $T_c(L)$ 
by the condition  $\langle W^2 \rangle = 0.516(1)$. Second, utilize 
the fact that the finite-size correction $\Delta T_c = T_c(L) \! - \! T_c(\infty)$ vanishes as $1/L^{1/\nu + \omega}$, with $\nu=0.6717$ being 
the 3D XY correlation length exponent \cite{universal3DXY}, to extrapolate $T_c(L)$ to the thermodynamic limit. This scheme is illustrated in Fig.~\ref{Fig:T_c_L_3D}. Note that the straight-line behavior of $\Delta T_c $ as a function of $1/L^{1/\nu + \omega}$ is yet another validation of the 3D XY criticalitity.

\begin{figure}[t]
\includegraphics[width=0.5\textwidth]{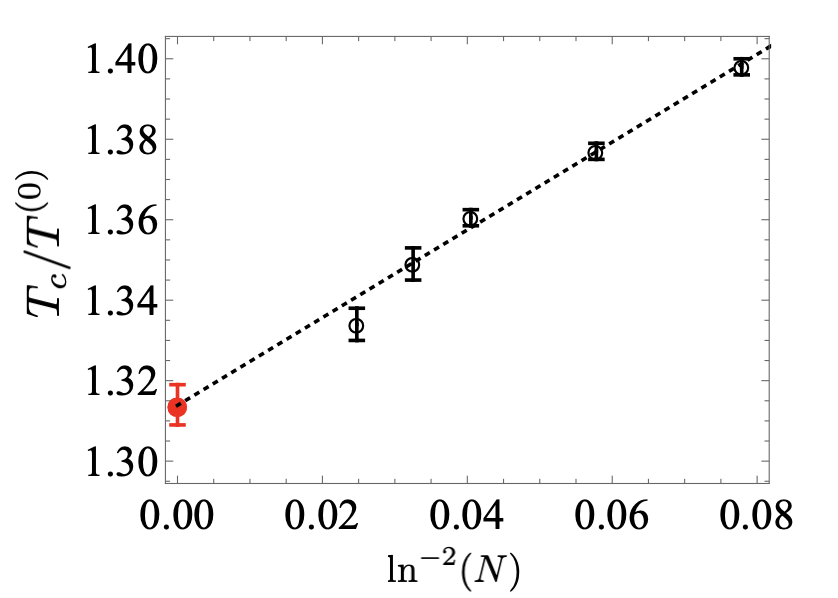}
\caption{Finite-size scaling of the ``critical'' temperature $T_c(N)$ for the 2D continuous system at $r_s=1.0$. The simulations were performed in a square cell with $N=36,\; 64, \; 144, \; 256$, and $576$ particles. 
After extrapolation to the $N \to \infty$ limit we obtain 
$T_c=1.314(5) T^{(0)}$ (red dot).}
\label{Fig:T_c_N_2D}
\end{figure}

The 2D case is different because the BKT transition lacks scale invariance. However, its asymptotically exact critical theory \cite{K_T,Kosterlitz_1974,N_K} allows one to eliminate 
the leading finite-size correction and obtain accurate thermodynamic limit results. Specifically, for a given number of particles, $N$, we define the ``critical'' temperature $T_c(N)$ as the temperature at which the ratio $\Lambda_s/T = \langle W^2 \rangle $ equals to its universal Nelson--Kosterlitz value of $2/\pi$ \cite{N_K,Note2}, in 
complete analogy with the 3D case discussed above.
Kosterlitz--Thouless renormalization-group theory predicts that 
the leading finite-size correction to $\Delta T_c = T_c(N) \! - \! T_c(\infty)$ scales as $1/ \ln^2 N$. This law is used to
extrapolate $T_c(N)$ to the $N\to \infty$ limit, and automatically validate the applicability of the BKT theory by the observation of  this (very specific) finite-size effect. The corresponding protocol is illustrated in Fig.~\ref{Fig:T_c_N_2D}.

%%%%%%%%%%%%%%%%%%%%%%%%%%%%%%%%%%%%%%%%% 
%%%%%%%%%%%%%%%%%%%%%%%%%%%%%%%%%%%%%%%%%%%%%%%%%

Figures \ref{Fig:T_c_3D} and \ref{Fig:T_c_2D} show final results for $T_c$, in 3D and 2D, respectively, as a function of $r_s$ defined by Eq.~(\ref{r_s}). 
In 2D, on approaching the Wigner crystal phase, which is estimated 
to emerge first at $r_s >50$, the phase diagram
is determined by competition between the Wigner crystal, hexatic,
superconductor, normal liquid, and an infinite set of emulsion phases \cite{Clark2009,emulsions}. Studies of this parameter regime go well beyond the scope of this Letter. 
In 3D, we restricted our study to $r_s \leq 50$, which corresponds
to densities all the way down to $na_B^3 \sim 10^{-5}$ (for greater values of $r_s$,
we start observing finite-T Wigner crystal states in accessible simulation cells). 

{\it Conclusions}.---Motivated by recent progress with quantifying 
scenarios of high-temperature bipolaron superconductivity, we studied 
the effect of Coulomb interactions on the critical temperature of the superconducting transition in the 2D and 3D Bose OCP, both on a lattice
and in continuous space. We provide evidence that 
the superconducting transition is in the XY universality class and 
obtain values for the critical temperature $T_c$ in a broad range of densities.

Screening of long-range interactions works in the direction of reducing their effect on $T_c$. However,
screening of $q\to 0$ modes involves divergent time scales (this is the prime reason for having plasmons instead of sound waves) making our finding that 
$T_c$ shifts are remarkably modest all the way to the Wigner crystal phase, surprising. 
The estimate (\ref{Tc_est}) remains quantitatively accurate and similar to that for systems with short-range interactions---up
to the reversal of the small and large density limits 
controlling the strength of interaction-induced correlations.  
The role of interactions is twofold: 
(i) On one hand (this aspect is conceptually important and well pronounced in 2D), interactions suppress thermal fluctuations of the superfluid order parameter amplitude (but not the phase), thereby contributing to the increase of $T_c$; 
(ii) On the other hand, strong local correlations caused by interactions 
increase phase fluctuations and suppress $T_c$. In a Coulomb system, the first effect is dominant at $r_s < 5$ (high-density limit) while the second effect 
takes over at $r_s > 10$ (low-density regime). The result of this 
competition is the $T_c/T_c^{(0)} $ {\it vs.} $r_s$ curve with a maximum at $r_s \sim 8$, 
see Figs.~\ref{Fig:T_c_3D} and \ref{Fig:T_c_2D}. 

We expect that our precise data for the dependence of $T_c$ on $r_s$ will prove important in the context of experimental realization of a relatively dilute, and, thus, well-defined bipolaron superconductor.

{\it Acknowledgments}.---NP and BS acknowledge support by the National Science Foundation under Grant No. DMR-2032077. MB acknowledges the support of the Natural Sciences and Engineering Research Council of Canada. CZ acknowledges support by the National Natural Science Foundation of China (NSFC) under Grant No. 12204173 and No. 12275263), the Innovation Program for Quantum Science and Technology (under Grant No. 2021ZD0301900), and the National Key R\&D Program of China (under Grants No. 2018YFA0306501).
% \bibliography{3DCoulombGas}

\begin{thebibliography}{99}
\bibitem{Note}  Equation (\ref{H_lattice}) does not
include coupling to the dynamic vector potential field. 
For the purpose of estimating $T_c$ in realistic systems 
with percent accuracy this weak relativistic effect is not relevant. It does change the universality class of the superconducting transition: In 3D, it converts the XY universality class into inverted XY; in 2D, it turns the BKT transition into a crossover (see, e.g., \cite{our_book}). These changes develop 
only in the close vicinity of $T_c$ when superconducting 
correlations are already set over large distances, and, thus, 
may not affect $T_c$ results derived on the basis of Eq.~(\ref{H_lattice}).
\end{thebibliography}

%%%%%%%%%%%%%%%%%%%%%%%%%%%%%%%%%%
\end{document}